\documentclass[twocolumn,aps,prl,showpacs,superscriptaddress]{revtex4}

\usepackage{graphicx}
\usepackage{amsmath}
\usepackage{amssymb}
\usepackage{latexsym}

\begin{document}

\title{Universality of state-independent violation of correlation inequalities\\for noncontextual theories}


\author{Piotr Badzi{\c a}g}
\email{pbg01@physto.se}
\affiliation{Department of Physics, Stockholm University, S-10691, Stockholm, Sweden}
\author{Ingemar Bengtsson}
\email{ingemar@physto.se}
\affiliation{Department of Physics, Stockholm University, S-10691, Stockholm, Sweden}
\author{Ad\'{a}n Cabello}
\email{adan@us.es}
\affiliation{Departamento de F\'{\i}sica Aplicada II, Universidad de Sevilla, E-41012
Sevilla, Spain}
\author{Itamar Pitowsky}
\email{itamarp@vms.huji.ac.il}
\affiliation{Department of Philosophy, The Hebrew University, Mount Scopus, Jerusalem
91905, Israel}


\date{\today}


\begin{abstract}
We show that the state-independent violation of inequalities for
noncontextual hidden variable theories introduced in [Phys. Rev.
Lett. \textbf{101}, 210401 (2008)] is universal, i.e., occurs for
any quantum mechanical system in which noncontextuality is
meaningful. We describe a method to obtain state-independent
violations for any system of dimension $d \ge 3$. This universality
proves that, according to quantum mechanics, there are no
``classical'' states.
\end{abstract}



\maketitle




\emph{Introduction.---}Bell inequalities \cite{Bell64} are
constraints involving the correlations of results of spacelike
separated measurements, which are satisfied by any local hidden
variable theory, but are violated by entangled states. For years,
entanglement has been considered ``\emph{the} characteristic trait
of quantum mechanics, the one that enforces its entire departure
from classical lines of thought'' \cite{Schrodinger35}.

Recently, one of us \cite{Cabello08} has shown that, for certain
physical systems, there are inequalities for the correlations of
compatible measurements which are satisfied by any noncontextual
hidden variable theory, but are violated by any quantum state, even
by nonentangled and totally mixed states. Specifically, Ref.
\cite{Cabello08} presents three correlation inequalities, two of
which are violated by any state described in quantum mechanics by a
Hilbert space of dimension $d=4$ (i.e., admitting 4 pairwise
compatible propositions), and a third which is violated by any state
of $d=2^N$ (with $N$ odd and $N \ge 3$).

An immediate question arises from this result: Can \emph{any}
quantum system be shown to violate an inequality which is valid for
any noncontextual hidden variable theory? Moreover, does this
violation hold for any state? The three inequalities in Ref.
\cite{Cabello08} are based on three special proofs of the Kochen and
Specker (KS) \cite{KS67, Mermin93} theorem in which each observable
appears in an even number of contexts, while the prediction of
quantum mechanics for the sums or the products of a compatible set
of these observables is minus the identity in an odd number of
contexts. A related question is therefore: Is there a method to
obtain correlation inequalities violated by any quantum state, based
on any available proof of the KS theorem, even those without this
special property? An affirmative answer would provide a method to
obtain state-independent violations for any $d \geq 3$, underlying
the universality of this phenomenon. This is the best result
possible, since two-dimensional quantum systems can be described by
contextual hidden variable theories \cite{KS67}.

The aim of this Letter is to give affirmative answers to these
questions. For any quantum system with $d \geq 3$ we find an
inequality which is satisfied by observables in any noncontextual
hidden variable theory, but violated by their corresponding quantum
observables, for any quantum state. We do that by obtaining an
inequality from a $d$-dimensional proof of KS theorem. These results
underline the universality of the phenomenon pointed out in
\cite{Cabello08} and allow us to draw the following conclusions: (i)
``classical'' states are impossible in quantum mechanics, and this
impossibility can be tested by experiments. (ii) In this
perspective, local realism underlying Bell inequalities can be
regarded as noncontextuality restricted to spacelike separated
contexts. Thus the inequalities derived here and Bell inequalities
belong to a larger family of inequalities satisfied by ``classical''
systems. Bell inequalities are usually optimized to allow for a
maximum violation on a particular quantum state, but here we seek
inequalities universally violated by all quantum states. The price
for this universality is a relatively small maximum degree of
violation allowed by quantum mechanics.

Some clarification of the terminology that will be subsequently used
might be in order. All the theories which we consider (quantum
mechanics, and trivially the noncontextual theories) satisfy the
principle of noncontextuality of probability. Suppose that
$\mathcal{A}$, $\mathcal{B}$, and $\mathcal{C}$ are physical
observables such that $\mathcal{A}$ is compatible with $\mathcal{B}$
and $\mathcal{C}$, but $\mathcal{B}$ is incompatible with
$\mathcal{C}$. The principle of noncontextuality of probability
states that, for every state, the expectation value of $\mathcal{A}$
is the same whether $\mathcal{A}$ is measured with $\mathcal{B}$, or
whether $\mathcal{A}$ is measured with $\mathcal{C}$. That is, the
expectation of an observable is context independent. In quantum
mechanics, noncontextuality of probability leads to Born's rule
(this is Gleason's theorem \cite{Gleason57}). Moreover, in quantum
mechanics, noncontextuality of probability implies nonsignaling. To
see this one can take $\mathcal{A}= \openone\otimes\mathcal{A}_{2}$,
$\mathcal{B}=\mathcal{B}_{1}\otimes\openone$, and
$\mathcal{C}=\mathcal{C}_{1}\otimes \openone$, with
$\mathcal{B}_{1}$ and $\mathcal{C}_{1}$ incompatible.

Another concept is the noncontextuality of values, which we simply
call noncontextuality. With every set of observables in classical
theories we can associate numerical values, which are within the
range taken by the observables, and respect the algebraic relations
among them. In quantum mechanics we cannot do that, and this is the
content of the KS theorem \cite{KS67, Mermin93}. Hence, hidden
variable theories which associate values with quantum mechanical
observables in the above manner must be contextual. The value of an
observable is context dependent.

Consider then a physical system admitting $d$ compatible dichotomic
observables (values $\pm 1$), hereafter denoted by $A_{1},\ldots,
A_{d}$, and consider $n$ different (mutually incompatible)
characterizations of this system via observable sets
$S^{j}=\{A_{1}^{j},\ldots, A_{d}^{j}\}$ with $j=1, \ldots, n$. The
different $S^{j}$'s will serve as different contexts. Some of the
mutually incompatible sets $S^{j}$ may overlap, so one can have
$j\neq k$, for which $A_{i}^{j}=A_{m}^{k}$ for some values of $i$
and $m$. The noncontextuality of probability implies that the
expectation of $A_{i}^{j}=A_{m}^{k}$ remains the same, whether
measured within the set (context) $S^{j}$ or $S^{k}$. As noted, the
situation is different when one tries to assign noncontextual values
to the observables. The KS theorem states that, for $d \ge 3$, there
are families of compatible dichotomic observable sets
$\{S^{j}\}_{j=1}^{n}$, such that it is impossible to consistently
assign noncontextual values $\pm 1$ to all the involved observables
$A_{i}^{j}$, so that exactly one observable in every set is assigned
one of the two values, e.g. $-1$ and the remaining observables are
assigned the other value, e.g. $+1$. Finding a suitable family of
sets $\{S^{j}\}_{j=1}^{n}$ establishes a proof of the KS theorem.

The question addressed here is whether one can derive from every
proof of the KS theorem an inequality for correlations, which will
be satisfied by every classical theory (i.e., whenever the
$A_{i}^{j}$ are interpreted as $\pm 1$ valued classical random
variables), and will be violated by the family $\{S^{j}\}_{j=1}^{n}$
for \emph{any} quantum state. Such an inequality can be tested in
the following way: Fix a quantum state, measure the observables in
the set $S^{1}$, then prepare another system in the same state and
repeat the experiment for $S^{2}$, and so on many times. Then, vary
the state and repeat it. A detailed description of a complete
experiment of this type is presented in \cite{ARBC09}. For the
procedure to make sense, the inequality must be written as a bound
on a function of components, each involving only compatible
observables.


\emph{Classical inequality.---}Consider a physical theory which
interprets the observables $A_{i}^{j}$ as classical random variables
with (simultaneous) noncontextual values $\pm 1$. We shall show that
it must satisfy the following inequality:
\begin{equation}
\beta (d,n)\leq n(d-2)-2, \label{inequality}
\end{equation}
where
\begin{equation}
\beta (d,n)=\sum_{j=1}^{n}\langle B^{j}\rangle,
\end{equation}
and
\begin{eqnarray}
B^{j} &=&-\sum_{p\neq q}A_p^{j} A_q^{j} - \sum_{p\neq q\neq r\neq
p}A_p^{j}A_q^{j}A_r^{j}-\cdots \notag \\
&&- \prod_{k=1}^{d}A_k^{j}-1.
\end{eqnarray}

The proof is as follows:
\begin{equation}
B^{j}=\sum_{k=1}^{d}A_k^{j}-\prod_{k=1}^{d}(1 + A_k^{j}).
\end{equation}
When $A_k^{j}=1$ for all $k$, then $B^{j}=d-2^{d}$. When, for at
least one value of $k$, $A_k^{j}=-1$, then
$B^{j}=\sum_{k=1}^{d}A_k^{j}$. For any $d > 2$, the former value is
smaller than all the latter values. Therefore, $B_{\max}^{j}=d-2$,
which is obtained for $d-1$ positive and one negative value of
$A_k^{j}$. Since the (overlapping) sets $S^{j}$ are chosen so that
it is impossible to produce $B_{\max }^{j}$ for all $j$ (because
$\{S^{j}\}_{j=1}^{n}$ yields a proof of the KS theorem), then an
upper bound for $\sum_{j=1}^{n}B^{j}$ is $n(d-2)-2$. Therefore, this
is also an upper bound for $\sum_{j=1}^{n}\langle B^{j}\rangle$.


\emph{Quantum violation.---}Quantum predictions violate inequality
(\ref{inequality}). If we associate every $A_k^{j}$ with a unit
vector $|v^{j,k}\rangle$ by
\begin{equation}
A_k^{j}=\openone - 2|v^{j,k}\rangle \langle v^{j,k}|,
\end{equation}
where $\langle v^{j,k}|v^{j,k^{\prime}}\rangle
=\delta_{kk^{\prime}}$ for every $1 \le j \le n$, then the operator
corresponding to the observable $B^{j}$ is
\begin{eqnarray}
B^{j} &=& \sum_{k=1}^{d} A_k^{j}-\prod_{k=1}^{d}(\openone + A_k^{j}) \\
&=&(d-2)\openone, \label{observable}
\end{eqnarray}
where equality (\ref{observable}) follows from the observation that
$\openone + A_k^{j}$ is twice the projection on the
$(d-1)$-dimensional subspace orthogonal to $|v^{j,k}\rangle$; hence
$\prod_{k=1}^{d}(\openone+A_{k}^{j})=0$ and thus (\ref{observable})
follows from $\sum_{k=1}^{d}A_k^{j}=(d-2)\openone$.

By summing Eq.~(\ref{observable}) over all the sets $S^{k}$, one
concludes that, independently of the quantum state, the results of
the measurements of the observables $B^{j}$ lead to a violation of
inequality (\ref{inequality}). Specifically, according to quantum
mechanics,
\begin{equation}
\beta _{\mathrm{QM}}=n(d-2). \label{quantumprediction}
\end{equation}


\emph{Impossibility of classical states.---}The affirmative answers
to our questions show that, for any quantum state, we can design an
experiment with an ensemble of particles in this state whose results
cannot be reproduced by any noncontextual hidden variable theory. In
this sense, all states of physical systems are nonclassical. A
totally mixed state is no exception. The measurements performed on
totally mixed states, if precise enough, will show their
nonclassicality all the same. There is always a finite separation
between a classical state and a quantum state. This difference can
be observed in actual experiments.


\emph{Bell inequalities are a particular case of more general
inequalities.---}Another consequence of the universality of the
state-independent violations of noncontextual inequalities is the
following. So far, on one hand, we had Bell inequalities derived
from the assumptions of local realism alone, and violated in a
state-dependent way by the quantum mechanical predictions. On the
other hand, we had proofs of the KS theorem which pointed out a
logical contradiction between the assumptions of noncontextual
realism and the formal structure of quantum mechanics---and for this
very reason, an experimental test of noncontextual realism is a
subtle matter indeed. Now, we see that Bell inequalities are, in a
sense, the tip of an iceberg. They belong to a more general family
of inequalities which are satisfied by appropriate classical random
variables, and violated by their corresponding quantum observables.
These include inequalities violated not only by entangled states of
composite quantum systems of , but for any state of any quantum
system with $d \geq 3$.


\emph{Remarks.---}Among all known proofs of the KS theorem for $d=3$
\cite{KS67, Schutte65, Belinfante73, PR85, CK90, Peres91}, the one
with the smallest $n$ has $n=36$ sets and $49$ observables
\cite{Schutte65}. Among all known proofs of the KS theorem for $d=4$
\cite{Peres91, ZP93, Kernaghan94, CEG96, Penrose00, AL98}, the one
with the smallest $n$ has $n=9$ sets and only $18$ observables
\cite{CEG96}. There are also proofs for other values of $d$
\cite{KP95, CEG05}, and methods to generate proofs of the KS theorem
for any value of $d$ \cite{ZP93, CEG05, CG96, PMMM05}.

For some $\{S^{j}\}_{j=1}^{n}$, the upper bound of inequality (\ref%
{inequality}) cannot be reached. For instance, for the $24$
observables in $d=4$ of \cite{Peres91}, $n=24$. The value of $\beta$
predicted by quantum mechanics is indeed $48$ [cf.
Eq.~(\ref{quantumprediction})], but in this case the upper bound for
$\beta(4,24)$ is $40$. It is substantially smaller than the general
bound in (\ref{inequality}) which is $46$. Therefore, the quantum
violation is in this case $8$, i.e., larger than our universal $2$.
This is due to the fact that the set of $24$ observables in
\cite{Peres91} is not critical (i.e., the proof also works if some
observables are removed) in the sense that it can generate $96$
(critical) $20$-observable and $16$ (critical) $18$-observable
proofs of the KS theorem \cite{CEG96}.

The case $d=4$ using the $18$ observables in \cite{CEG96} deserves a closer
examination. The resulting inequality contains a sum of $99$ terms bounded
by $16$, while the quantum prediction is $18$. However, if we omit all the
correlations but those between 4 observables, then we obtain a $9$-term
inequality introduced in \cite{Cabello08}. The bound there is $7$ while
quantum mechanics allows $9$. All this shows that the method presented here
may lead to inequalities that are not optimal in the sense that they may be
strictly weaker than the inequalities with fewer terms but the same
violation. Finding simpler inequalities with the same violation is
interesting since in actual experiments every expectation value is affected
by errors.

Our inequality (\ref{inequality}) is related to earlier ``KS
inequalities'' \cite{SBZ01, Larsson02} between probabilities instead
of correlations. Although an equivalence can be established between
the final inequalities, the main difference is that, while the
derivation of the inequalities in \cite{SBZ01, Larsson02} assumes
the sum rule (i.e., it requires quantum mechanics), the derivation
of inequality (\ref{inequality}) \emph{only} requires the assumption
of noncontextual probabilities (i.e., it does not require quantum
mechanics). Quantum mechanics is only used to predict that
(\ref{inequality}) will be violated by the experimental results.


\emph{Conclusions.---}To sum it up, we have produced an algorithm
associating an inequality for the results of compatible measurements
with every proof of the KS theorem. The inequality is satisfied by
any noncontextual hidden variable theory. Nevertheless, it is
violated by quantum mechanical predictions for every physical state,
including the seemingly classical totally mixed state. In this
sense, our result shows that there is no such thing as a classical
state, and suggests that Bell inequalities are a particular type of
a more general inequalities where neither spacelike separation nor
entanglement play a fundamental role.


\begin{acknowledgments}
The authors thank J.-\AA . Larsson for valuable discussions. A.\,C.
thanks the Department of Physics of Stockholm University for its
hospitality, and acknowledges support from projects
No.~P06-FQM-02243 and No.~FIS2008-05596. I.\,P. acknowledges the
support of the Israel Science Foundation Grant No.~744/07.
\end{acknowledgments}



\begin{thebibliography}{99}


\bibitem{Bell64}
 J.~S. Bell,
 Physics (Long Island City, N.Y.) \textbf{1}, 195 (1964).


\bibitem{Schrodinger35}
 E. Schr\"{o}dinger,
 Proc. Cambridge Philos. Soc. \textbf{31}, 555 (1935).


\bibitem{Cabello08}
 A. Cabello,
 Phys. Rev. Lett. \textbf{101}, 210401 (2008).


\bibitem{KS67}
 S.~Kochen and E.~P.~Specker,
 J. Math. Mech. \textbf{17}, 59 (1967).

\bibitem{Mermin93}
 N.~D.~Mermin,
 Rev. Mod. Phys. \textbf{65}, 803 (1993).


\bibitem{Gleason57}
 A.~M.~Gleason,
 J. Math. Mech. \textbf{6}, 885 (1957).

\bibitem{ARBC09}
 E. Amselem, M. R{\aa }dmark, M. Bourennane, and A. Cabello
 (to be published).


\bibitem{Schutte65}
 K.~Sch\"{u}tte in 1965, first reported by K.~Svozil in 1994.
 See also J.~Bub,
 Found. Phys. \textbf{26}, 787 (1996).

\bibitem{Belinfante73}
 F.~J. Belinfante,
 \emph{A Survey of Hidden-Variables Theories}
 (Pergamon Press, New York, 1973).
 See also E.~de Obaldia, A.~Shimony, and F.~Wittel,
 Found. Phys. \textbf{18}, 1013 (1988).

\bibitem{PR85}
 A. Peres and A. Ron,
 in \emph{Microphysical Reality and Quantum Formalism},
 edited by A.~van der Merwe, F.~Selleri, and G.~Tarozzi
 (Kluwer, Dordrecht, 1988), Vol.~2, p.~115.

\bibitem{CK90}
 J.~H. Conway and S. Kochen around 1990,
 first reported in A. Peres,
 \emph{Quantum Theory: Concepts and Methods}
 (Kluwer, Dordrecht, 1993), p.~114.
 See also J.~H.~Conway and S.~Kochen,
 in \emph{Quantum [Un]speakables: From Bell to Quantum Information},
 edited by R.~A. Bertlmann and A.~Zeilinger
 (Springer-Verlag, Berlin, 2002), p.~257.

\bibitem{Peres91}
 A. Peres,
 J. Phys. A \textbf{24}, L175 (1991).


\bibitem{ZP93}
 J.~R. Zimba and R.~Penrose,
 Studies in History and Philosophy of Science \textbf{24}, 697 (1993).

\bibitem{Kernaghan94}
 M. Kernaghan,
 J. Phys. A \textbf{27}, L829 (1994).

\bibitem{CEG96}
 A. Cabello, J.~M. Estebaranz, and G. Garc\'{\i}a-Alcaine,
 Phys. Lett. A \textbf{212}, 183 (1996).

\bibitem{Penrose00}
 R. Penrose,
 in \emph{Quantum Reflections},
 edited by J.~Ellis and D.~Amati
 (Cambridge University Press, Cambridge, England, 2000), p.~1.

\bibitem{AL98}
 P.~K.~Aravind, and F.~Lee-Elkin,
 J. Phys. A \textbf{31}, 9829 (1998).


\bibitem{KP95}
 M. Kernaghan and A. Peres,
 Phys. Lett.~A \textbf{198}, 1 (1995).

\bibitem{CEG05}
 A.~Cabello, J.~M.~Estebaranz, and G.~Garc\'{\i}a-Alcaine,
 Phys. Lett. A \textbf{339}, 425 (2005).


\bibitem{CG96}
 A. Cabello and G.~Garc\'{\i}a-Alcaine,
 J. Phys. A \textbf{29}, 1025 (1996).

\bibitem{PMMM05}
 M. Pavi\v{c}i\'{c}, J.-P. Merlet, B.~D. McKay, and N.~D. Megill,
 J. Phys.~A \textbf{38}, 1577 (2005).


\bibitem{SBZ01}
 C. Simon, \v{C}. Brukner, and A. Zeilinger,
 Phys. Rev. Lett. \textbf{86}, 4427 (2001).

\bibitem{Larsson02}
 J.-\AA . Larsson,
 Europhys. Lett. \textbf{58}, 799 (2002).


\end{thebibliography}
\end{document}